\documentclass{tlp}
\usepackage{color}
\usepackage{latexsym}

\title[Worst-Case Groundness Analysis using $\Def$]
      {Worst-Case Groundness Analysis using \\
       Definite Boolean Functions}
\author[Samir Genaim, Jacob M. Howe and Michael Codish]
       {SAMIR GENAIM\\
         Department of Computer Science\\
         Ben-Gurion University
       \and JACOB M. HOWE\\
         Computing Laboratory\\
         University of Kent
       \and  MICHAEL CODISH\\
         Department of Computer Science\\
         Ben-Gurion University}
\newcommand{\Pos}{\textsf{Pos}}
\newcommand{\Def}{\textsf{Def}}
\newtheorem{proposition}{Proposition}[section]
\newtheorem{theorem}{Theorem}[section]

\begin{document}

\maketitle
\thispagestyle{empty}
\begin{abstract}
  This note illustrates theoretical worst-case scenarios for
  groundness analyses obtained through abstract interpretation over
  the abstract domains of \emph{definite} ($\Def$) and \emph{positive}
  ($\Pos$) Boolean functions.
  For $\Def$, an example is given for which any $\Def$-based abstract
  interpretation for groundness analysis follows a chain which is
  exponential in the number of argument positions as well as in the
  number of clauses but sub-exponential in the size of the program.
  For $\Pos$, we strengthen a previous result by illustrating an
  example for which any $\Pos$-based abstract interpretation for
  groundness analysis follows a chain which is exponential in the size
  of the program.
 It remains an open problem to determine if the worst case for $\Def$
 is really as bad as that for $\Pos$.
\end{abstract}

\section{Introduction}

Boolean functions play an important role in various formal methods for
specification, verification and analysis of software systems.  In
program analysis, Boolean functions are often used to approximate
properties of the set of states encountered at a given program point.
For example, a conjunction $x\wedge y$ could specify that variables
$x$ and $y$ satisfy some property whenever control reaches a given
program point. A Boolean function $\varphi_1 \rightarrow \varphi_2$
could specify that if $\varphi_1$ is satisfied at a program point
(perhaps depending on the unknown inputs to the program) then also
$\varphi_2$ is satisfied. A disjunction $\varphi_1 \vee\varphi_2$
could arise as a consequence of a branch in the control where
$\varphi_1$ and $\varphi_2$ approximate properties of the
\texttt{then} and \texttt{else} branches respectively.

For program analysis using Boolean functions, we often consider the
\emph{positive} Boolean functions, $\Pos$. Namely, those for which
$f(1,\ldots,1) = 1$ (denoting $false$ and $true$ by $0$ and $1$
respectively).  This restriction is natural as, due to the element of
approximation, the result of an analysis is not a ``\emph{yes/no}''
answer, but rather a ``\emph{yes/maybe not}'' answer. In this case
there is no ``negative'' information. Sophisticated $\Pos$-based
analyzers implemented using binary decision diagrams
\cite{Bryant:CS92} have been shown \cite{CLCVH95} to give good
experimental results with regards to precision as well as the
efficiency of the analyzers. However, scalability is a problem and
inputs (programs) for which the analysis requires an exponential
number of iterations or exponentially large data structures are
encountered \cite{codish-worstcase}.

The domain, $\Def$, of definite Boolean functions is a subdomain of
$\Pos$. These are the positive functions whose sets of models are
closed under intersection. The domain $\Def$ is less expressive than
$\Pos$. For example, the formula $x\lor y$ is not in $\Def$. However,
$\Def$-based analyzers can be implemented using less complex data
structures and can be faster than $\Pos$-based analyzers.
For goal dependent groundness analyses (where a description of the
inputs to the program being analyzed is given) $\Def$ has been shown
to provide a reasonable tradeoff between efficiency and precision
\cite{King:esop99, King:esop00}.

The work described in \cite{codish-worstcase} illustrates a series of
pathological inputs for $\Pos$-based groundness analysis. That paper
defines a predicate $chain(x_1,\ldots,x_n)$ using $n$ clauses and
illustrates that its $\Pos$-based groundness analysis requires $2^{n}$
iterations. However, given that the size of the program (the total
number of arguments), is quadratic in $n$ ($m=n^2+n$), the
number of iterations is sub-exponential in the size of the input
($2^{n}$ or $2^{O(\sqrt{m})}$).
It has been suggested that $\Def$ analyses might provide better
scalability properties than $\Pos$ due to the restriction to functions
whose models are closed under intersection.
This note makes two contributions:
\begin{enumerate}
\item It demonstrates that the worst-case behavior of a $\Def$-based
  analysis is (at least) as bad as that described in
  \cite{codish-worstcase} for $\Pos$-based analyses; and
\item It demonstrates that the worst-case behavior of a $\Pos$-based
  analysis is exponential in the size of the input.
\end{enumerate}
We have not succeeded to demonstrate a worst-case analysis for $\Def$
for which the number of iterations is exponential in the size of the
input, nor to prove that $\Def$-based groundness analysis has
sub-exponential worst-case behaviour. This remains an open problem.

\section{A potential worst-case for $\Def$}\label{s2}

Consider an $n$-ary Boolean function $f$. A model $M$ of $f$ can be
viewed as a sequence $(b_1,\ldots,b_n)$ of zero's and one's such that
$f(b_1,\ldots,b_n)=1$. 
For the sake of our construction, we order $n$-ary models according to
their value as $n$-digit binary numbers. So a model $M_1$ is smaller
or equal to a model $M_2$ if and only if the binary number
corresponding to $M_1$ is less or equal to the binary number
corresponding to $M_2$.
The intersection of models is defined as usual so that
$(a_1,\ldots,a_n) \cap (b_1,\ldots,b_n) = (c_1,\ldots,c_n)$ where
$c_i=1$ if and only if $a_i=b_i=1$.

Let us first comment on the series of programs which demonstrates the
potential worst-case behavior of a $\Pos$-based groundness analysis
\cite{codish-worstcase}. The analysis of the predicate $chain/n$
enumerates the models of the (constant) $n$-ary Boolean function $1$
($true$) in reverse order. Starting from the initial approximation
(which has no models), each consecutive approximation is a function
which has one new model that was not in the previous iteration. For
example, when $n=3$, the models accumulate in the following order:
$(1,1,1),(1,1,0), (1,0,1), (1,0,0),\ldots,(0,0,0)$ and the
$\Pos$-based analysis totals 8 iterations.
In contrast the corresponding $\Def$-based analysis totals 4
iterations because at each iteration the current set of models is
closed under intersection. So for example, in the third iteration, the
set $\{(1,1,1),(1,1,0), (1,0,1)\}$ is closed to give
$\{(1,1,1),(1,1,0), (1,0,1),(1,0,0)\}$.

We now construct a series of programs which demonstrates the potential
worst-case behavior of a $\Def$-based groundness analysis. This
construction is based on the following observation:

\begin{proposition}\label{prop1}
Let $M$ be an $n$-ary model. Then the set of $n$-ary models
smaller or equal to $M$ is closed under intersection.
\end{proposition}

\begin{proof}
  The result follows from the following observation: If $M_1$ and
  $M_2$ are $n$-ary models, then $M_1\cap M_2$ is no larger than $M_1$
  (and no larger than $M_2$). This is because $M_1\cap M_2$ is
  obtained from $M_1$ (or from $M_2$) by changing some one's to
  zero's. 
\end{proof}

A consequence of Proposition \ref{prop1} is that the domain of
definite Boolean functions over $n$ variables contains a chain of
length $2^n$. To demonstrate such a chain consider an enumeration
$M_0,\ldots,M_{2^n-1}$ of the $n$-ary models according to their binary
ordering (so $M_0=(0,\ldots,0)$ and $M_{2^n-1}=(1,\ldots,1)$). Observe
that $M_i$ is the $n$-ary binary representation of $i$.  Define a
sequence $F=(f_0,\ldots,f_{2^n-1})$ as follows: let $f_0$ be the
Boolean function with the empty set of models and for $0<i<2^n-1$
define $f_i$ to be the Boolean function whose models are
$\{M_0,\ldots,M_{i-1}\} \cup \{M_{2^n-1}\}$.
>From the construction it is clear that $F$ forms a chain. Moreover,
the elements of $F$ are in $\Def$: They are positive because they have
$M_{2^n-1}$ as a model; and from Proposition \ref{prop1}, it follows
that they are closed under intersection. The chain $F$ is of length
$2^n-1$ because, for $1<i<2^n-1$ $f_i$ has exactly one model more than
$f_{i-1}$.
This is the setting for our construction.

The ($\Def$-based) groundness analysis of the following predicate
``$p/n$'' iterates through the chain $F$. The arguments typeset in
boldface highlight the case for $n=4$.  The program size is quadratic
in $n$ and consists of a single predicate of arity $n$ with $n+1$
binary clauses. The analysis of the program can be viewed as counting
from zero to $2^n-2$ in its arguments.

{\small
\definecolor{light}{gray}{0.25}
\def\m#1{ \mbox{\textcolor{light}{$#1$}}}
\[\begin{array}{lcrclcl}
p(\m{X_n},    &\m{\ldots,}& \bf X_4, X_3, X_2, c )&\leftarrow& 
              p(\m{X_n,}   &\m{\ldots,}& \bf X_4, X_3, X_2, X_1).\\
p(\m{X_{n-1},}&\m{\ldots,}& \bf X_3, X_2, c , X_1)&\leftarrow& 
              p(\m{X_{n-1},} &\m{\ldots,}& \bf X_3, X_2, X_1, c ).\\
p(\m{X_{n-2},}&\m{\ldots,}& \bf X_2, c , X_1, X_1)&\leftarrow& 
              p(\m{X_{n-2},} &\m{\ldots,}& \bf X_2, X_1, c , c ).\\
&\vdots\\
p(\bf c  , X_1,&\m{\ldots,}& \bf X_1, X_1)&\leftarrow& 
              p(\bf X_1, c , &\m{\ldots,}& \bf c , c ).\\
p(\bf X_1, X_1,&\m{\ldots,}& \bf X_1, X_1)&\hspace{-8mm}.&
\end{array}\]
}



\section{A Challenge}

The $\Def$- and $\Pos$-based groundness analyses of the predicate
$p/n$ program in the series given in this note involve an exponential
number of iterations and compute an $n$-ary Boolean function. The same
is true for the $\Pos$-based analysis of the series given in
\cite{codish-worstcase}.
However, it is important to note that complexity is typically
expressed in terms of the size of the input to a problem and that the
size of the program defining $p/n$ in both series is quadratic in $n$
($m=n^2+n$). Hence, formally speaking, we have shown that both $\Def$
and $\Pos$-based groundness analyses may potentially involve a number
of iterations which is $2^{O(\sqrt{m})}$.  This is bad enough, but
sub-exponential.

For $\Pos$, we can strengthen the result. The following program is of
size linear in $n$ ($m=11\cdot n$) and its $\Pos$-based groundness
analysis requires $2^n-2$ iterations.

{\small
\[\begin{array}{l}
\begin{array}{l}
p(X_1, \ldots, X_1). \\
p(A_1, \ldots, A_n) ~\leftarrow~
         p(B_1, \ldots, B_n), ~s(A_1, \ldots, A_n,~B_1, \ldots, B_n).\\
\end{array} \\[2ex]
\begin{array}{llcll}
s(c~, X_1, \ldots, X_1, & X_1, c~, \ldots,  c~ ).\\
s(W, A_1, \ldots,A_{n-1}, & W, B_1, \ldots,B_{n-1}) & \leftarrow &
         s(A_1,\ldots,A_n, & B_1,\ldots,B_n).\\
\end{array}
\end{array}
\]
}

Intuitively, the $2n$ arguments of the predicate $s/2n$ represent two
$n$-digit binary numbers (the first is the successor of the second) so
that the $n$ recursive clauses from the program in Section \ref{s2}
can be simulated by two clauses for $s/2n$. The base case of $s/2n$
corresponds to the last recursive clause. However, the analysis of
$s/2n$ does not follow an exponential chain so we still need the
predicate $p/n$ to get the worst-case behaviour.
This approach does not work for $\Def$ because the result in $\Pos$
for $s/2n$ is not closed under intersection.

\section{Conclusion}

We have demonstrated a $2^{O(m)}$ worst case complexity for $\Pos$ and
at least $2^{O(\sqrt{m})}$ for $\Def$ (where $m$ is the size of the
program). It remains to be determined if the worst case for $\Def$ is
really as bad as that for $\Pos$ or perhaps $\Def$ has better
worst-case behaviour.

\begin{theorem}
Groundness analysis using \Def\ has a potentional
worst-case behaviour involving $2^{O(\sqrt{m})}$ iterations, where $m$
is the size of the program.    
\end{theorem}

\begin{theorem}
Groundness analysis using \Pos\ has a
  worst-case behaviour involving $2^{O(m)}$ iterations, where $m$ is
  the size of the program.  
\end{theorem}

\bibliographystyle{tlp}


\end{document}